# Bibliometric Perspectives on Medical Innovation

# using the Medical Subject Headings (MeSH) of PubMed

Loet Leydesdorff,[1] Daniele Rotolo,[2] and Ismael Rafols [3,4]

**Abstract**

Multiple perspectives on the nonlinear processes of medical innovations can be distinguished and combined using the Medical Subject Headings (MeSH) of the Medline database. Focusing on three main branches—"diseases," "drugs and chemicals," and "techniques and equipment"—we use base maps and overlay techniques to investigate the translations and interactions and thus to gain a bibliometric perspective on the dynamics of medical innovations. To this end, we first analyze the Medline database, the MeSH index tree, and the various options for a static mapping from different perspectives and at different levels of aggregation. Following a specific innovation (RNA interference) over time, the notion of a trajectory which leaves a signature in the database is elaborated. Can the detailed index terms describing the dynamics of research be used to predict the diffusion dynamics of research results? Possibilities are specified for further integration between the Medline database, on the one hand, and the *Science Citation Index* and *Scopus* (containing citation information), on the other.

[1] Amsterdam School of Communication Research (ASCoR), University of Amsterdam, Kloveniersburgwal 48, 1012 CX Amsterdam, The Netherlands; loet@leydesdorff.net; http://www.leydesdorff.net.
[2] SPRU (Science and Technology Policy Research), University of Sussex, Freeman Centre, Falmer Brighton, East Sussex BN1 9QE, United Kingdom; d.rotolo@sussex.ac.uk.
[3] SPRU (Science and Technology Policy Research), University of Sussex, Freeman Centre, Falmer Brighton, East Sussex BN1 9QE, United Kingdom; i.rafols@sussex.ac.uk.
[4] INGENIO (CSIC-UPV), Universitat Politècnica de València, València, Spain.

## 1. Introduction

A bibliometric perspective on innovations and inventions is difficult to obtain because innovations by definition occur across scientific, technological, and economic domains that are archived using different databases and classifications, and hence from different perspectives. Whereas bibliometrics has focused on output indicators of the science & technology system such as publications and patents, economists can consider patents and other knowledge carriers as input to "total factor productivity" (TFP; Solow, 1957). As Grilliches (1994, at p. 14) noted, "our current statistical structure is badly split, there is no central direction, and the funding is heavily politicized."

Patent literature, for example, is organized in databases other than scientific literature, although cross-references have been organized systematically (e.g., in the Derwent Innovation Index) and also exploited in empirical studies (Boyack & Klavans, 2011; Glänzel & Meyer, 2003; Grupp, 1996; Narin & Noma, 1985; Narin & Olivastro, 1992). However, the classifications and codifications in these databases have remained very different from those in scientific literature. For example, citations may mean something different for patents than for scholarly texts.

Furthermore, given the non-linear nature of innovation (Nelson & Winter, 1977, 1982), feedback loops between different stages and perspectives can be expected to prevail (Kline & Rosenberg, 1985), and therefore the information may evolve as in a whirl. In a methodological appendix to their evaluation of the cancer mission in the USA during the 1970s, Studer & Chubin (1980, at p. 269) asked: "Does it make sense to attempt to 'control' for one relationship while studying others? What would be meant by ‘controling for ideas’ or ‘controling for cocitations?’ If disparate dimensions of science are not carefully analyzed in their own terms, the possibility of relating their respective contributions is nil.” For the study of knowledge-based innovations, one would need to be able to move from representations of contexts of discovery to contexts of application, and *vice versa* (Gibbons *et al*., 1994), in order to map path-dependencies, yet without losing control of understanding how the interacting systems are further developed, both recursively and in relation to one another.

Which are the relative weights of the different channels within and between the databases? Within each of the databases, the problem of developing baselines for the measurement of contributions and change have been addressed, and much progress has been made during the past decade. Klavans & Boyack (2009), for example, noted an increasing consensus in mapping the structure of science using multidisciplinary databases such as the Web-of-Science (WoS) or Scopus. Within this approach, interactive overlays have been developed which allow researchers to position their samples using journals as indicators of intellectual organization (e.g., Leydesdorff & Rafols, 2012; Moya-Anegón *et al*., 2007; Rafols et al., 2010; Rosvall & Bergstrom, 2009). Small & Garfield (1985) pointed already to the geography of nations as a second baseline for the comparison. Interactive overlays of Google Maps (possible since approximately 2007) enable us to map both publications and patents in terms of institutional addresses (Bornmann & Leydesdorff, 2011; Leydesdorff & Persson, 2010). The interactive mapping of patents in terms of International Patent Classifications (IPC) seems nowadays within reach (Newman *et al*., 2011; Schoen *et al*., 2011; cf. Leydesdorff & Bornmann, in press).

Disciplines and technologies vary in terms of the distances between the sciences and their applications in the economy (Narin & Olivastro, 1992). Patents are indicators of inventions; a main function of patents is legal protection of intellectual property rights, but they can also be considered as input indicators to the economy or output indicators of science given a linear model. The non-linear interactions, however, crosscut the intellectual organization at levels of aggregation lower than journals or patent classes, that is, at the level of specific documents.

Agarwal & Searls (2008 and 2009) proposed using the documents in the Medline database and their different classifications as a potential source of information on innovation drivers that can be mined as literature. Following up on this suggestion, one can develop different perspectives on the Medline database. Would this possibility to switch perspectives provide us with new insights in the interactive process of medical innovations? Is one able to construct an overlay which informs us about the dynamics of innovation perhaps in an early stage by focusing on the socio-cognitive interactions between practitioners and users in the clinic, on the one hand, and laboratory scientists, on the other?

Whereas Agarwal & Searls (2009) conceptualized the innovative interactions in terms of "demand" or "need" (represented as "diseases") versus "supply" (represented as new "drugs and chemicals"), we prefer a nonlinear model in which these classification systems are considered as different perspectives on the same data. A third branch of the index referring to "Analytical, Diagnostic and Therapeutic Techniques and Equipment" (hereafter "Techniques and Equipment") can provide yet another perspective relevant to medical innovations (Von Hippel, 1976, 1988; Blume, 1992).

The nonlinearity in the interactions among the perspectives makes it necessary to use visualization and animation techniques for the recognition and specification of possible patterns. The static analysis provides us with a geometrical metaphor that is predictably insufficient when contexts are also changing. The analyst's mental map can be overburdened when both the latent structures and the observable variables change. Using various baselines with the same data may enable us to consider complementary perspectives on the complex dynamics in which the structural dynamics can be recognized as clusters represented and potentially moving in multi-dimensional spaces.

## 2. The Medline database

MeSH terms are provided on the basis of intensive indexing articles of more than 5,500 leading journals for the Medline/PubMed databases, a service of the National Institute of Health (NIH), the largest funder of biomedical research in the world. The investment of NIH in maintaining the National Library of Medicine (NLM) is on the order of US$ 100M (Hicks & Wang, 2011, at p. 292).[5] Unlike the Web-of-Science (WoS) or Scopus, Medline does not cover the full range of disciplines; but a large part of the scholarly literature in the life sciences is included even more exhaustively than in the comprehensive databases (Lundberg *et al.*, 2006).

[5] At http://www.nlm.nih.gov/about/2011CJ.html#budget_auth , NLM formulates its Budget Policy as follows: "The FY 2011 Budget request is $118.843 million, an increase of $3.776 million or 3.3 percent from the FY 2010 appropriation of $115.067 million. In FY 2011, the Library will concentrate on maintaining its current level of services and enhancing and expanding some of its most heavily used resources, including Medline/PubMed and PubMed Central, which provide critical access to published biomedical research results worldwide. Another key service, MedlinePlus, contains a wide range of information written and formatted for consumers. Keeping MedlinePlus current with new information (in English, Spanish and other languages) from NIH and other reliable sources is a high priority in FY 2011. NLM will support the expansion of ClinicalTrials.gov in FY 2011 to accommodate the reporting provisions of the Food and Drug Administration Amendments Act of 2007."

The classification at the article level is elaborated in great detail, with a hierarchical tree covering 16 separate branches that can reach up to twelve levels of depth (Table 1). Unlike other disciplinarily specialized databases such as Chemical Abstracts (Bornmann *et al.*, 2009), the multiple tree-structure of the *Index Medicus* allows for mapping documents differently across heterogeneous domains. For the aim of this paper—to develop a tool for the investigation of the innovation process in the medical sector from multiple perspectives—we focus on three main branches arguably relevant to the process of medical innovation: (*i*) “Diseases” (category C), (*ii*) “Drugs & Chemicals” (category D), and (*iii*) “Techniques and Equipments” (category E).

**Table 1**: MeSH tree branches and levels.

| **Branch** | **MeSH Terms** | **Levels** |
|---|---|---|
| A – Anatomy | 2798 | 11 |
| B – Organisms | 5136 | 12 |
| C – Diseases | 10923 | 10 |
| D – Chemicals and Drugs | 19933 | 11 |
| E – Analytical, Diagnostic and Therapeutic Techniques and Equipment | 4105 | 10 |
| F – Psychiatry and Psychology | 1078 | 7 |
| G – Phenomena and Processes | 3138 | 10 |
| H – Disciplines and Occupations | 476 | 6 |
| I – Anthropology, Education, Sociology and Social Phenomena | 603 | 9 |
| J – Technology, Industry, Agriculture | 558 | 10 |
| K – Humanities | 213 | 6 |
| L – Information Science | 481 | 9 |
| M – Named Groups | 231 | 7 |
| N – Health Care | 2191 | 8 |
| V – Publication Characteristics | 182 | 4 |
| Z – Geographicals | 497 | 7 |

*Note. The full MeSH tree structure can be found at http://www.nlm.nih.gov/mesh/trees.html.*

The Medline database—which is the constitutive core of the PubMed interface[6] at http://www.ncbi.nlm.nih.gov/pubmed/advanced—is also integrated into the Web-of-Knowledge (WoK), which contains the WoS allowing for citation searching. López-Illescas *et al*. (2009), for example, used MeSH categories to enrich the journal sets under study across WoS and Scopus data. Bordons *et al.* (2004) successfully tracked the medical applications of a drug (aspirin) using MeSH categories. Perianes-Rodríguez *et al.* (2011) noted that many pharmaceutical corporations have relatively similar profiles in journal-based maps. For example, the difference between Pfizer and Merck could not be captured in terms of their publications in medical journals. Bornmann *et al*. (2008) suggested that MeSH terms might be more appropriate classifiers than journals or journal groups for the measurement of research performance and impact.

Can a representation of the translation processes between contexts of discovery and application be obtained by mapping the practices of medical communities as represented in the Medline data (cf. Blume, 1992)? Could this type of integration in terms of perspectives across boundaries between representations that have been organized for different audiences provide us with a perspective on mapping and perhaps supporting innovation processes with bibliometric means? Can specific patterns in innovation processes thus be retrieved and specified?

[6] MEDLINE is the largest component of PubMed (http://pubmed.gov/), the freely accessible online database of biomedical journal citations and abstracts created by the U.S. National Library of Medicine (NLM®). Approximately 5,400 journals published in the United States and more than 80 other countries have been selected and are currently indexed for MEDLINE. A distinctive feature of MEDLINE is that the records are indexed with NLM's controlled vocabulary, the Medical Subject Headings (MeSH®). Source: http://www.nlm.nih.gov/pubs/factsheets/dif_med_pub.html.

## 3. The project design

We focus on the public interface of PubMed and use the MeSH categories for both the static mapping of scholarly literature and, thereafter, the dynamic animation of a single science-based innovation trajectory. Building on previous mapping efforts (e.g., Rafols *et al.*, 2010), we aim to develop and test the software for a potential upscale in the same pass. The long-term perspective is to map innovation trajectories across different contexts (Akera, 2007). However, the research question of this study is first limited to the possibilities and limitations of mapping innovations in terms of Medline documents and MeSH categories.

With the objective of visualization in mind, we limited the analysis to the first and second level of the index, containing 117 first-level and 1664 second-level categories. For example, "Cardiovascular diseases" is a first-level category among the diseases (with tree number C14), whereas C14.260 is a second-level category for "Cardiovascular infections". ("Bacterial endocarditis" would be a third-level category; C14.260.249.)[7] For the exploration, we focused first on "Diseases" as a single category ("C") among the 16 main branches of the index. (The full MeSH tree structure can be found at http://www.nlm.nih.gov/mesh/trees.html.)[8] Thereafter, we upscaled to the full set, but the results were too complex for an appreciation. Therefore, we limited ourselves to the three branches—"C," "D," and "E"—that we considered as most relevant for the study of medical innovations. Dedicated software was developed which allows users to overlay document samples on base maps for these three branches. The base maps are organized into files which can be read into Pajek[9] or VOSviewer.[10] The Pajek-format is often used as a medium of interchange among different (freeware) visualization programs. As against spring-embedded algorithms, the MDS-like visualizations of VOSviewer offer some conceptual advantages (see Leydesdorff & Rafols, 2012). The files generated by VOSviewer allow the user additionally to webstart visualizations at the internet (Van Eck & Waltman, 2011).

[7] The system is further complicated because "Cardiovascular infections" is not only the second-level category C14.260, but also the third-level category C01.539.190 under the second-level category of "Infection" and with a subcategory "C01.539.190.249 for "Endocarditis, Bacterial." However, the downloaded data (at PubMed) is organized in terms of the labels and categories with similar names can therefore not be easily distinguished in this data. We did not attempt to correct for this.

[8] See also at http://en.wikipedia.org/wiki/List_of_MeSH_codes.

The routine for generating these overlay files from PubMed data is available at http://www.leydesdorff.net/pubmed/pubmed.exe. One also needs a spreadsheet with the base map information (at http://www.leydesdorff.net/pubmed/pubmed.dbf.) The user is first asked to download a document set relevant for his/her research at the PubMed interface (as a result of an advanced search) in the so-called tagged format of Medline. Alternatively, one can use the Medline search option in the Web-of-Knowledge (WoK) and download the results in the tagged format. The program first prompts for a choice between these two sources, and then generates:

1. A matrix that contains the documents of the sample as the row variables and all MeSH categories attributed to these records as column variables. This file can, for example, be read into SPSS (e.g., for factor analysis). A file named "labels.sps" in the SPSS syntax is made available containing the MeSH categories as variable descriptors;
2. A file "pajek.vec" which can be used as a vector given the base map of second-level MeSH terms—available at http://www.leydesdorff.net/pubmed/pubmed.paj—in order to generate a visual overlay using Pajek (Rafols *et al.*, 2010; cf. De Nooy *et al.*, 2005);

[9] Pajek is a freeware program for the analysis and visualization of large networks available at http://pajek.imfm.si/doku.php?id=download.
[10] VOSviewer is a (freeware) visualization program available at http://www.VOSviewer.com.

3. A file “vos.txt” which can be opened similarly in VOSviewer for the generation of an overlay; this overlay can also be brought online for a webstart (Van Eck & Waltman, 2011).

The user should be aware that the output files are overwritten in a next run and thus need to be saved in a separate folder for future use (if so wished). More detailed instructions can be found at http://www.leydesdorff.net/pubmed/index.htm.

## 4. Methods and materials

### *4.1 The generation of a base map from the PubMed data*

The data for a base map was retrieved using the search string ("2010"[Publication Date] : "2010"[Publication Date]) at the advanced search interface of PubMed (at http://www.ncbi.nlm.nih.gov/pubmed/advanced). The year 2010 was the latest full year available at the time of the downloads (October/November 2011). The above-mentioned search string recalled 923,086 records on October 27, 2011, including all publications entered into this database with a publication date within the year 2010. Within this initial sample, 922,786 records (> 99.9%) could be downloaded with 9,634,935 attributions of MeSH categories and 5,999,426 of subcategories (qualifiers).[11]

We first constructed the grand matrix of these records as cases versus the 1,781 unique MeSH terms at the first and second level of the index tree as column variables. (Specific tree-categories can be selected from this larger set by setting filters to the database processing, and sub-matrices can be constructed accordingly.) The matrices are factor-analyzed in SPSS v.19 using Varimax rotation and (*i*) the SPSS default extraction of the number of factors with eigenvalues larger than unity, and (*ii*) a ten-factor solution. This information provides us with insights into the network structure among the categories in the respective analyses.

[11] This set of documents was composed by 4,562,911 (co-)authors.

Cosine-normalized matrices were generated (from the various document/MeSH matrices; cf. Ahlgren *et al*., 2003) within SPSS and then used as input to Pajek (version 2.01), for both analysis and visualization. The cosine file can be saved in Pajek as a ".net" file that can also be read into VOSviewer (version 1.4.0) or other visualization software.

### *4.2 Options for the visualization*

Visualization programs (Pajek, Gephi, etc.) may have a problem with the cluttering of the labels in the case of a large number of variables (Leydesdorff *et al*., 2011). The developers of VOSviewer solved this problem elegantly by foregrounding and backgrounding labels in accordance to their prominence in the network (Van Eck & Waltman, 2010). Additionally, the use of the MDS-like optimization in VOSviewer can be considered as an advantage: MDS minimizes stress in the representation in terms of the multivariate system (using Kruskall's (1964) stress function or a similar measure), whereas graph-analytically inspired algorithms—for example, spring-embedders such as Fruchterman & Reingold (1991) and Kamada & Kawai (1989)—optimize in terms of individual linkages in the network (Leydesdorff & Rafols, 2012).

One possible disadvantage of this suppression of overlapping labels in VOSviewer might be, in the case of MeSH, that labels of the most relevant categories in a specific search can be so close to one another that some of them could inadvertently become invisible in the maps. Using spring-embedded algorithms, however, Pajek allows flexibly for zooming, manual relocations, and the extraction of subnetworks, whereas the maps of VOSviewer are fixed. We shall discuss the pros and cons of using both overlays below, but leave the eventual choice to the user. Our routines provide both options because the quality and relevance of the two visualizations may differ from case to case depending on the research question and the subject matter under study.

The base maps are stored in the files "pubmed.paj" for Pajek and "pubmed.dbf" for VOSviewer, respectively. Both files are needed for the layout of a map; they can be retrieved at http://www.leydesdorff.net/pubmed/index.htm. We first tested the system using the search results for "Opthof[Author]" which generated a recall of 153 records (on November 12, 2011) because this author—specialized in cardiovascular diseases at the Academic Medical Center of the University of Amsterdam—has a unique author name and collaborated with one of us in bibliometric projects. Opthof provided us with feedback that led to error-correction and further fine-tuning of the overlay maps. The files were thereafter upscaled for the case of RNA interference (RNAi; see Section 6 below).

## 5. Results

### *5.1. Initial comparison of sets and subsets*

Table 2 provides the descriptive statistics when using (*i*) only the main branch "C" of the MeSH tree for diseases, (*ii*) the three main branches "C," "D," and "E" most relevant for processes of medical innovation, and (*iii*) all 16 branches. Only 590,257 of the 922,786 documents retrieved (63.4%) contain MeSH categories at this relatively high level of the tree. The other documents include also data imported into the database without being part of Medline or being owned by the National Library of Medicine (NLM).[12]

**Table 2**: Descriptive statistics of the three sets compared.

| MeSH Branche(s) | Number of Records | MeSH Categories | MeSH used in 2010 | Largest Component |
|---|---|---|---|---|
| "C" | 106,337 | 332 | 303 | 288 |
| "C", "D", and "E' | 324,357 | 895 | 839 | 822 |
| All 16 branches | 590,257 | 1,781 | 1,534 | 1,520 |

*Notes. The MeSH tree structure in 2011 is adopted for the analysis.*
*The number of MeSH categories refers to the first and the second-level of MeSH tree.*
*The largest component is evaluated for cosine > 0.01.*

Table 3 shows the results of the factor-analytic exploration and informs us, for example, that almost half of the factors have an eigenvalue larger than unity. These factors explain only 50-55% of the variance in the matrices. The first ten factors still explain less than 5% in all three cases. In other words, the factor structure is very weak: the co-variation in the links among the categories plays a minor role in comparison to the percentage of variance explained by the variables themselves.

[12] Only papers with both the status ("STAT –") of "MEDLINE" and the ("OWN –") field labeled as "NLM"—the abbreviation for the National Library of Medicine—are provided with MeSH. The journal *Scientometrics*, for example, is owned by the NLM, but not included in Medline. Therefore, this journal is available in the PubMed database, but without MeSH terms.

**Table 3**: Results of the factor analyses of matrices of different subsets.

| %Variance explained<br>Second-level categories of: | eigenvalue > 1<br>(a) | factors/ variables<br>(b) | 10 factors<br>(c) |
|---|---|---|---|
| "C" | 53.65 | (150/303) = 49.50% | 4.42 |
| "C", "D", and "E' | 51.28 | (393/839) = 47.35% | 1.81 |
| All 16 branches | 50.70 | (698/1534) = 45.50% | 1.12 |

In summary, there is almost no co-variation at the second level of the index.[13] Within a single category (e.g., "Diseases") one can find some network structure, but using more main categories, one finds hardly any structure. Leydesdorff (2008) found a similar effect in the case of the International Patent Classification (IPC) of the World Intellectual Property Organization (WIPO). Agarwal & Searls (2009, at p. 869), however, reported considerable networking at lower levels of aggregation using only disease-related MeSH terms.

[13] Do the 16 branches of the tree stand almost orthogonal to one another? At the first level of the index 117 MeSH terms are available of which 76 are used in 55,379 documents in 2010. The factor structure is a bit more pronounced: 63.88 % of the variance can be explained by 46 factors with an eigenvalue > 1, and 15.39% can be explained by the ten main factors. However, overwhelmingly the factor loadings are between -0.1 and 0.1. Thus, there is virtually no factor structure at this level.

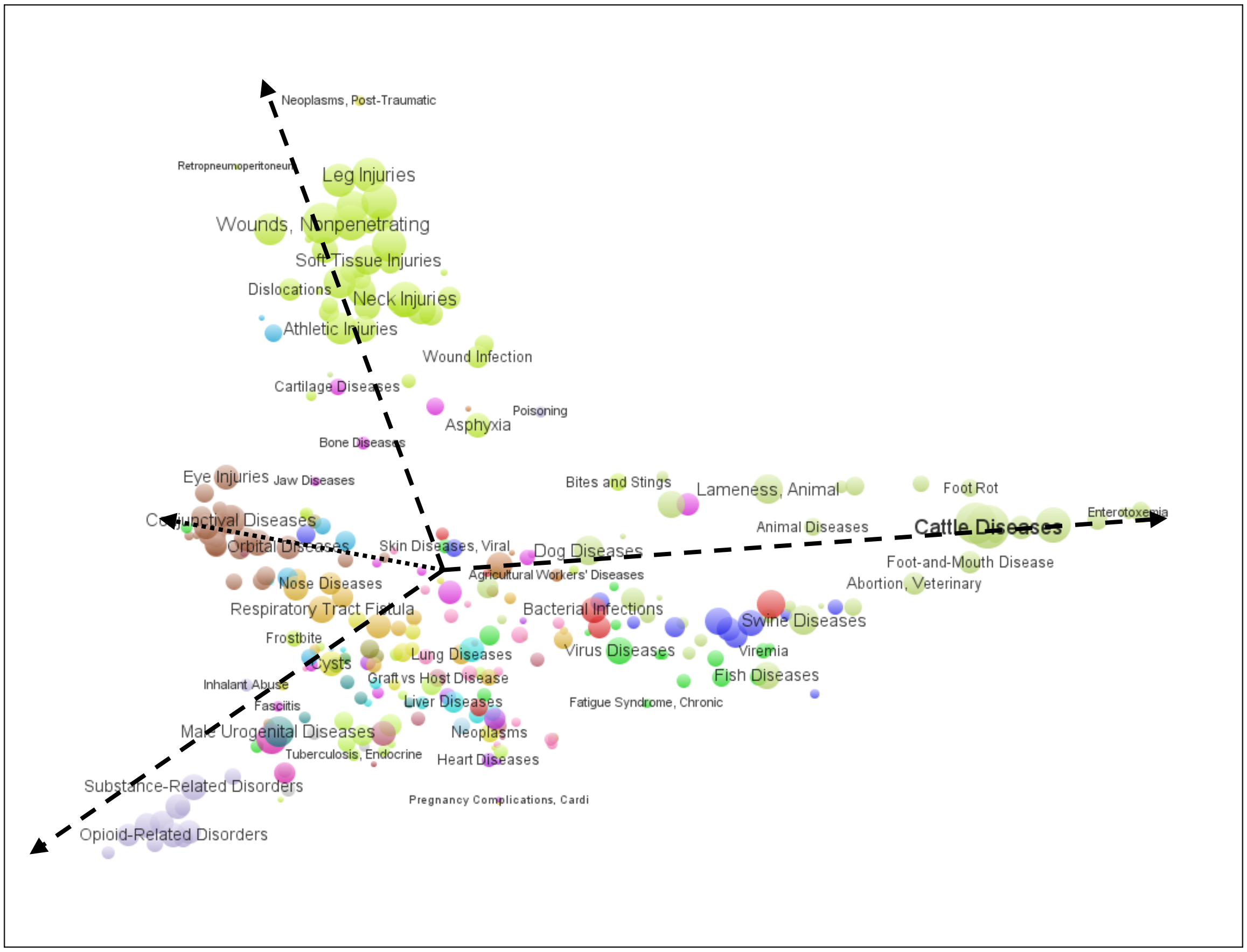

**Figure 1:** 288 MeSH categories at the second-level of the branch "C", diseases; cosine > 0.01; $N$ of documents = 106,337. Sizes are proportionate to [$\log_2(n+1)$];[14] colors were automatically generated by the clustering algorithm in VOSviewer (cf. Leydesdorff & Rafols, 2012).

Figure 1 provides the visualization using the 288 MeSH categories in the largest component of the network of categories for "Diseases." This representation shows three (or perhaps four) main axes (added by us):[15] a major axis to the right about animal diseases; an axis to the left at the top about wounds and diseases of the extremities; an axis to the left-bottom about addiction and perhaps narcotics; and a cluster to the left-middle about infections in the respiratory tract.

---

[14] The size of the nodes is proportional with the $\log_2(n+1)$ in order to prevent single occurrences ($n$ = 1) from disappearing (because the log(1) = 0).

[15] The colors are automatically generated by an algorithm in VOSviewer for the decomposition (Waltmann et al., 2011). However, they can be changed on the basis of other classifications as discussed by Leydesdorff & Rafols (2012).

When all 16 main branches are combined in a single visualization (Figure 2), the multi-dimensionality of the set can no longer be sufficiently reduced to visualization in two dimensions. For example, different world regions are positioned in the central area. (This can be considered as a projection of an orthogonal axis into this plane.) These other branches, however, are not so interesting from our perspective of medical innovation studies. For this reason, we decided to focus only on three perspectives for the mapping of innovations.

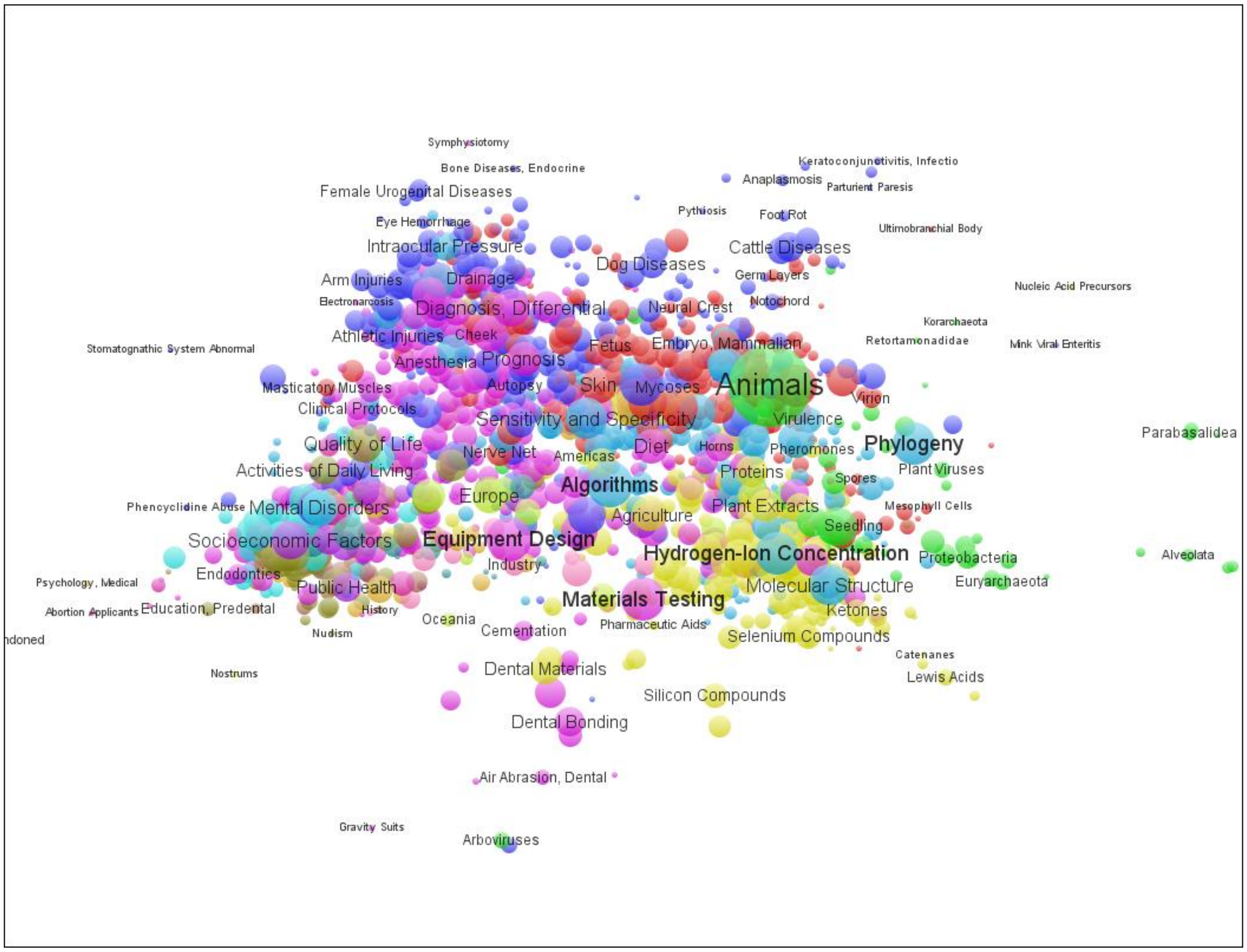

**Figure 2:** Largest component (enlarged) of 1,520 MeSH at the second level used in 2010. (Colors accord with the 16 main branches of the tree; cosine > 0.01.)

*5.2. The base map for innovation trajectories*

Both Pajek and VOSviewer provided us with interesting and comparable base maps when the 822 MeSH in the three branches (C: "Diseases;" D: "Drugs & Chemicals;" and E: "Techniques and Equipment") were used (Figure 3). Of the 895 MeSH terms in the subset of these three main branches of the index, 839 were attributed to 324,357 records in 2010; 822 of these categories formed a largest component at a threshold level of cosine > 0.01.[16]

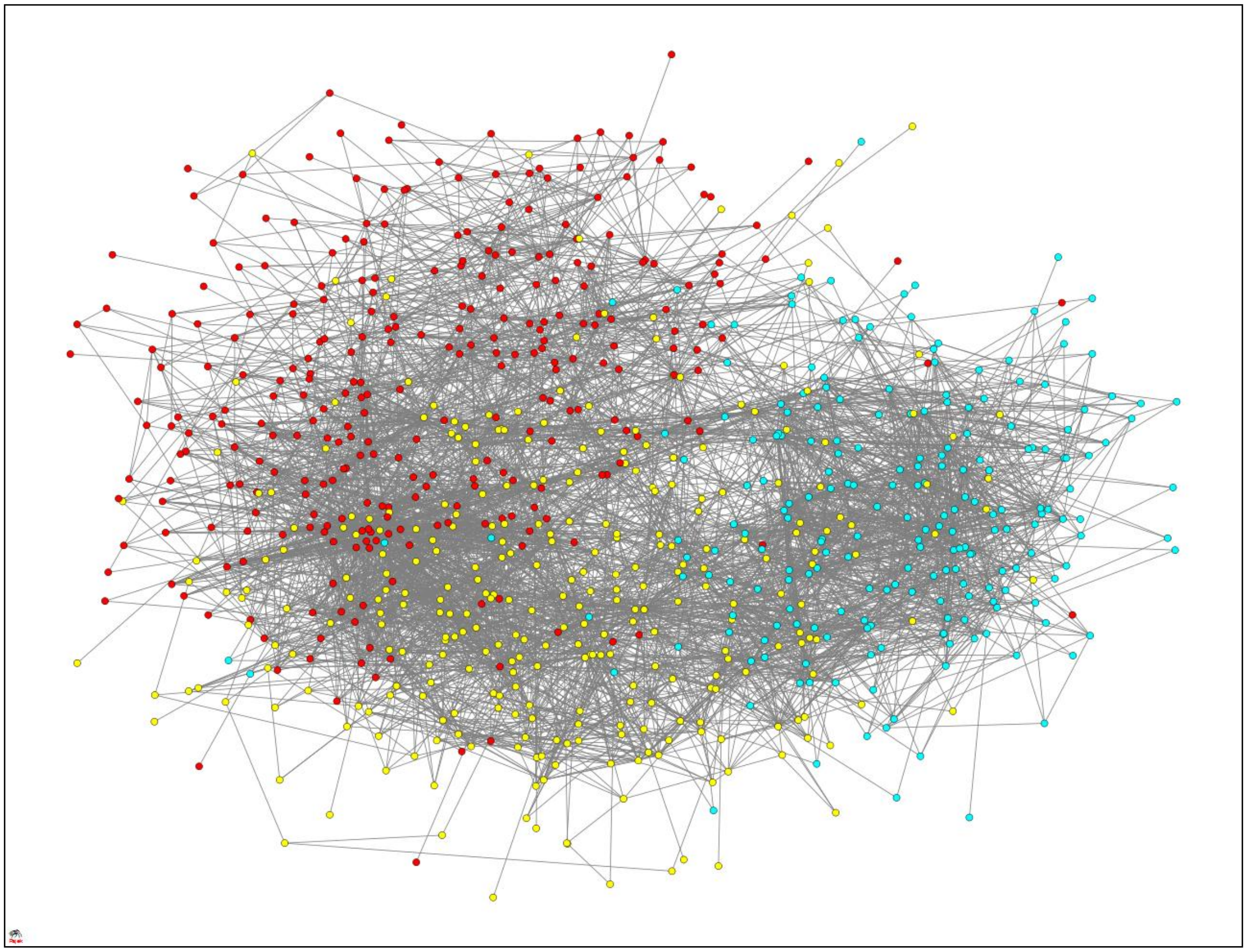

**Figure 3**: 822 MeSH (*N* of documents = 324,357); "Diseases" red, "Drugs & Chemicals" light blue, "Techniques & Equipment" yellow; cosine > 0.01.

[16] A low-level threshold is needed for the visualization because marginal cosine values in the third or higher decimals prevail among all categories and thus the network would be too densely packed for the purpose of visualization. Values of cosine ≤ 0.01 were set equal to zero; otherwise the file remained unchanged.

Figure 3 shows this base map in Pajek using three colors: red for “Diseases”, light blue for “Drugs & Chemicals”, and yellow for “Techniques and Equipment”. As expected, the three sets do not overlap strongly, although the cluster of nodes representing “Techniques and Equipment” permeates into the two neighbouring domains more than the two others between them.

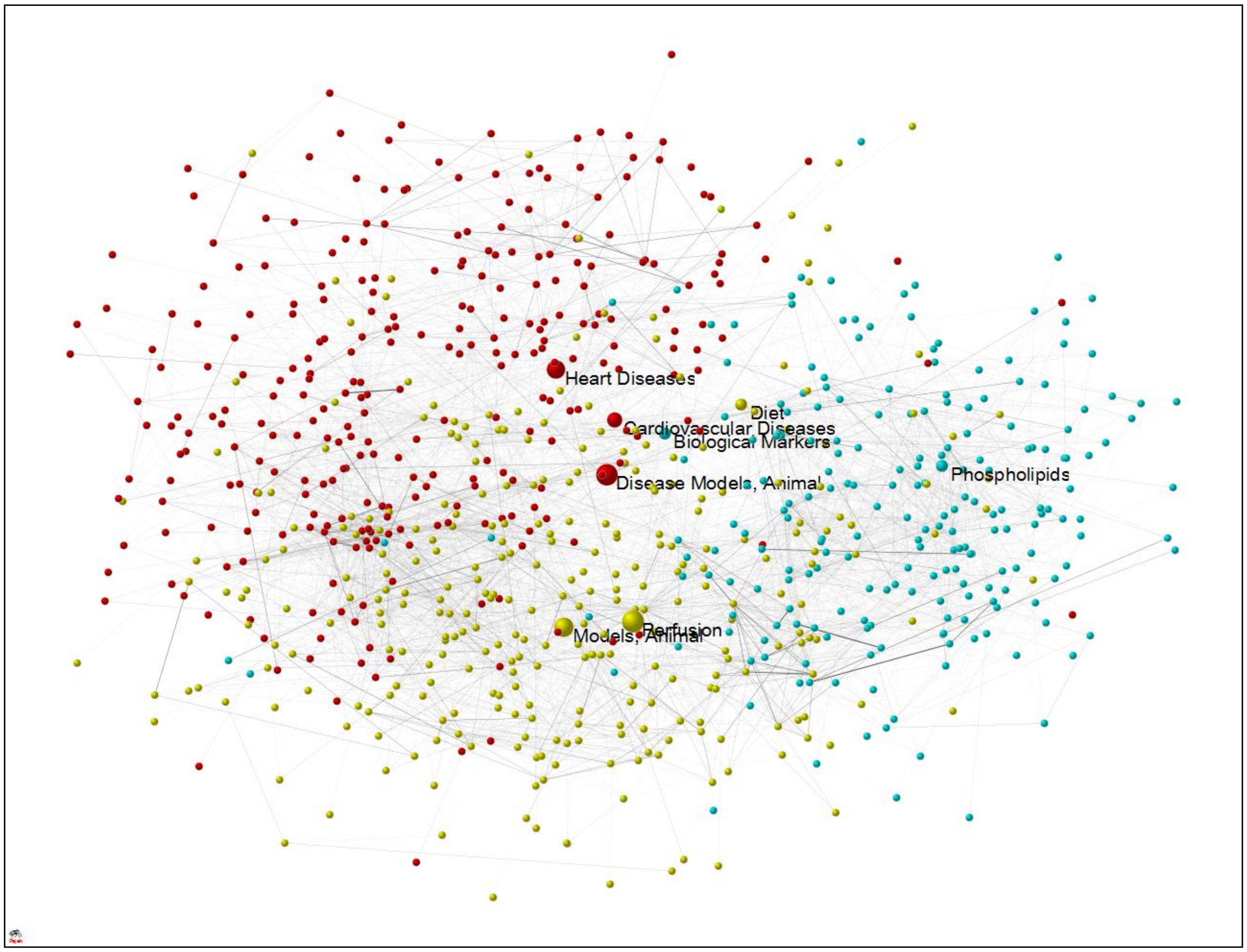

**Figure 4**: 8 MeSH categories at the second level (occurring 31 times) and attributed to 153 documents (coauthored by Opthof) in the PubMed database (cosine > 0.01; the size is proportionate to the $\log_2$ of the number of occurrences plus one).[14]

Figure 4 shows the overlay of 153 documents (co-)authored by Opthof. We provide this map primarily for methodological reasons; that is, in order to compare it with the equivalent overlay in VOSviewer in terms of the quality of the visualization (Figure 5).

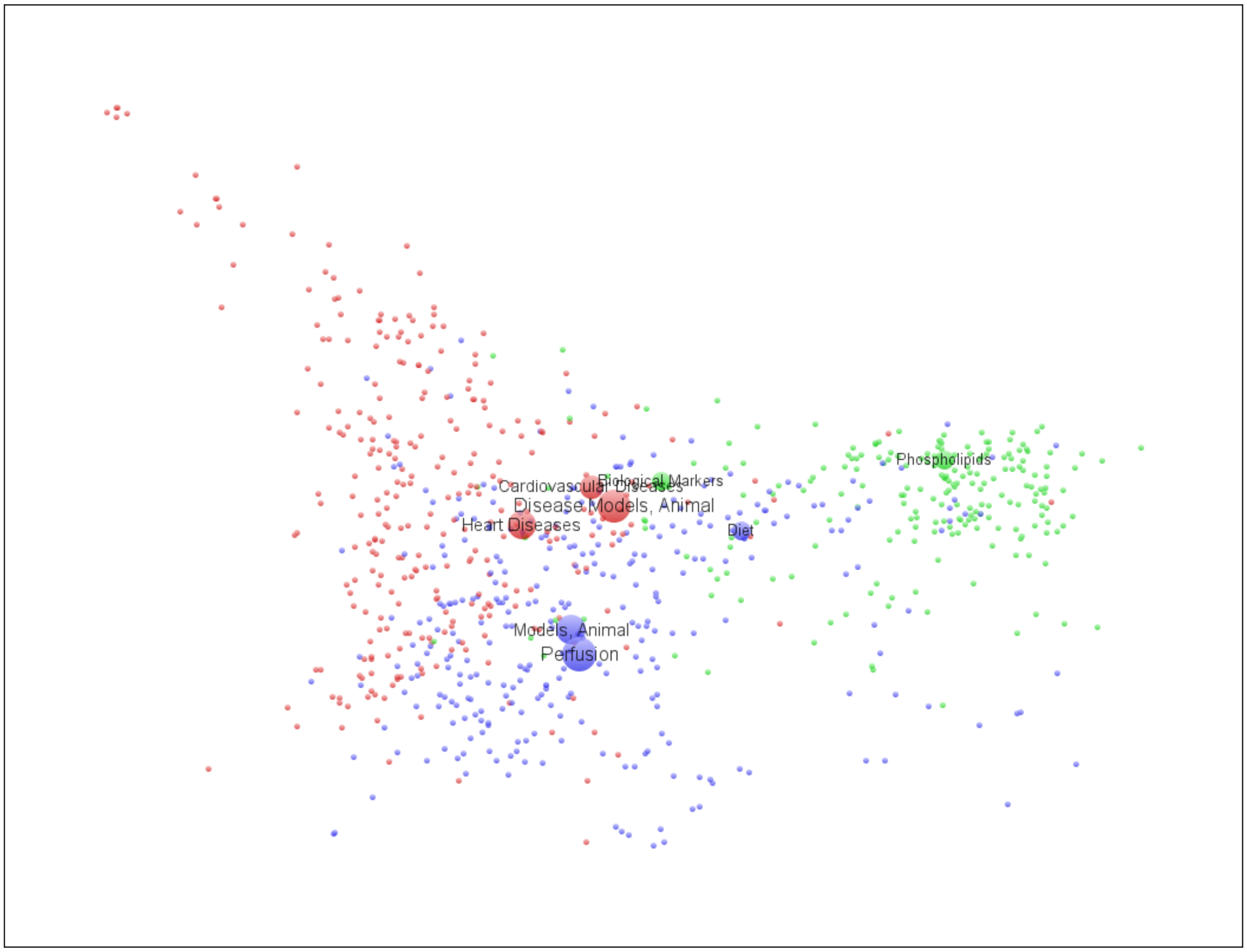

**Figure 5**: 8 MeSH categories at the second level (occurring 31 times) and attributed to 153 documents (coauthored by Opthof) in the PubMed database (cosine > 0.01; the size is proportionate to the $\log_2$ of the number of occurrences (plus one);[14] normalized weights).

The comparison between the Figures 4 and 5 enables us to specify the advantages and disadavantages of the two representations. However, let us first note that the two representations are not essentially different in structure or dimensionality. In both cases, one finds the "Diseases" red-colored to the upper-left side, and the "Drugs & Chemicals" to the right. The latter are blue-colored in Pajek (Figure 4) and green-colored using VOSviewer (Figure 5). (The user can change colors in both these programs.) The third category of "Techniques & Equipment" is dominant in the lower parts of the maps, but permeates more into the two neighbouring areas than these latter two permeate into each other's domains. Thus, the maps enable us to see the position of document sets in these three dimensions both spatially and in terms of the coloring attributed. As noted, the size of the nodes in both maps is proportionate to the logarithm of the occurrence frequencies.[14]

In VOSviewer, one can choose between overlapping and non-overlapping labels. We shall use the non-overlapping labels in the examples below, but here (in Figure 5) we use the map *with overlapping labels* in order to show the problem that the overlap can be even denser using VOSviewer when compared with Pajek. Whereas in the MDS-like solution of VOSviewer the central categories are often positioned in the central area, a spring-embedded algorithm such as Kamada & Kawai's (1989) tends to spread the groups of categories in the visualization. MDS (and VOSviewer), however, provide layouts that are spanned in terms of the latent dimensions of the network (as in factor analysis; Schiffman *et al.*, 1981). As analyzed above (Table 3), the factor structure was not pronounced in the case of this network.

## 6. The case of "RNA Interference"

RNA Interference (RNAi) is a gene-silencing technique discovered and further developed in basic science, but with a variety of clinical applications; for example, in cancer and genetic diseases (Sung & Hopkins, 2006). RNAi became topical after the publication of an article in *Nature* in 1998 entitled "Potent and specific genetic interference by double-stranded RNA in *Caenorhabditis elegans*" (Fire *et al*., 1998). The two principal investigators—Craig Mello at the University of Massachusetts Medical School in Worcester, and Andrew Fire at Stanford University—received the Nobel Prize in medicine for this breakthrough in 2006.

RNAi research focused initially on the use of exogenous small RNA molecules with potentially large effects in the suppression of gene expression by inhibiting the messenger RNA. These molecules have become known as siRNA (that is, small-interfering RNA). More recently, it was discovered that RNA can also endogenously generate the same mechanism of messenger RNA suppression. These endogenous RNA have become known as microRNA (miRNA).

Leydesdorff & Rafols (2011) studied the emergence of the technology associated with "small interfering RNA" ("siRNA") in terms of papers in the *Science Citation Index*; Leydesdorff & Bornmann (in press) mapped patents with a similar search string in the databases of the US Patent and Trademark Office (USPTO) as an overlay to Google Maps. Using both patents and patent applications, Lundin (2011) suggested that the problem of the delivery of the molecules at the places of intervention in the body were not yet solved, and therefore the clinical research stagnates. Leydesdorff & Bornmann (in press) confirmed the predominance of reagent suppliers instead of drug-developing corporations in the US patent portfolios in this field.

Can the perspective of including medical applications as provided by the PubMed database provide us with additional insights into this development? Three MeSH terms have in the meantime been included that classify the new technology: "RNA interference" (index tree number: G05.355.315.203.374.790), "MicroRNAs" (D13.150.650.319), and "Small Interfering

RNA" (D13.150.650.700). Several other MeSH terms can also be associated with the new technology, but less specifically so. Using these MeSH terms for the retrieval, however, one would no longer be able to distinguish clearly between the emergence of the technology in terms of MeSH categories and the emergence of these categories themselves in the *Index Medicus*.[17] For this reason, we used an informed search string formulated as follows: "((((siRNA[Title/Abstract]) OR RNAi[Title/Abstract]) OR interference RNA[Title/Abstract]) OR RNA interference[Title/Abstract]) OR miRNA[Title/Abstract]) OR micro RNA[Title/Abstract]) OR interfering RNA[Title/Abstract])".

**Table 4**: Descriptive statistics of the recall with the search string AND publication dates.

| Year | Number of Records | Number of Records with MeSH Terms | Number of MeSH Terms | Number of Categories | MeSH Terms Active in the Overlay |
|---|---|---|---|---|---|
| 1997 | 9 | 9 | 108 | 77 | 0 |
| 1998 | 19 | 19 | 256 | 148 | 0 |
| 1999 | 35 | 35 | 546 | 242 | 3 |
| 2000 | 112 | 111 | 1,606 | 486 | 2 |
| 2001 | 172 | 171 | 2,544 | 655 | 5 |
| 2002 | 454 | 447 | 7,177 | 1,442 | 19 |
| 2003 | 1,197 | 1,189 | 19,175 | 2,599 | 42 |
| 2004 | 2,554 | 2,536 | 42,565 | 3,945 | 103 |
| 2005 | 3,955 | 3,924 | 64,881 | 4,935 | 113 |
| 2006 | 5,486 | 5,430 | 81,747 | 5,639 | 113 |
| 2007 | 6,745 | 6,663 | 99,888 | 6,385 | 151 |
| 2008 | 7,756 | 7,613 | 114,072 | 6,808 | 164 |
| 2009 | 8,542 | 8,327 | 124,645 | 7,113 | 171 |
| 2010 | 9,816 | 9,303 | 137,937 | 7,586 | 207 |
| Total | 46,852 | 45,777 | 697,147 | 48,060 | 1,093 |

[17] For example, only nine records were retrieved with these three MeSH terms in 2000, whereas the search string recalled 111 records. In 2010, however, 11,164 records were retrieved with the three MeSH search terms, against 9,816 with the search string, but the overlap was only 6,521. Thus, these two approaches can lead to considerably different retrievals.

Table 4 provides the descriptive statistics of the recalls in each year.[18] In 1998, for example, nineteen records were found with 256 MeSH terms attributed to 148 categories, but none of these categories were among the 822 MeSH terms used for the baseline map. Although the original article of Fire *et al*. (1998) was among these 19 documents, it was attributed with MeSH terms such as "Gene Expression Regulation" at levels of the tree deeper than two, and not among the three main categories here under study (diseases, drugs, and techniques). "Gene Expression Regulation", for example, is indexed at level 3 (G05.355.315) and among "Phenomena and Processes" (in the G-branch of the index tree). The step towards "Drugs and Chemicals" was not yet visible in this early stage of the development at the second level of the index.

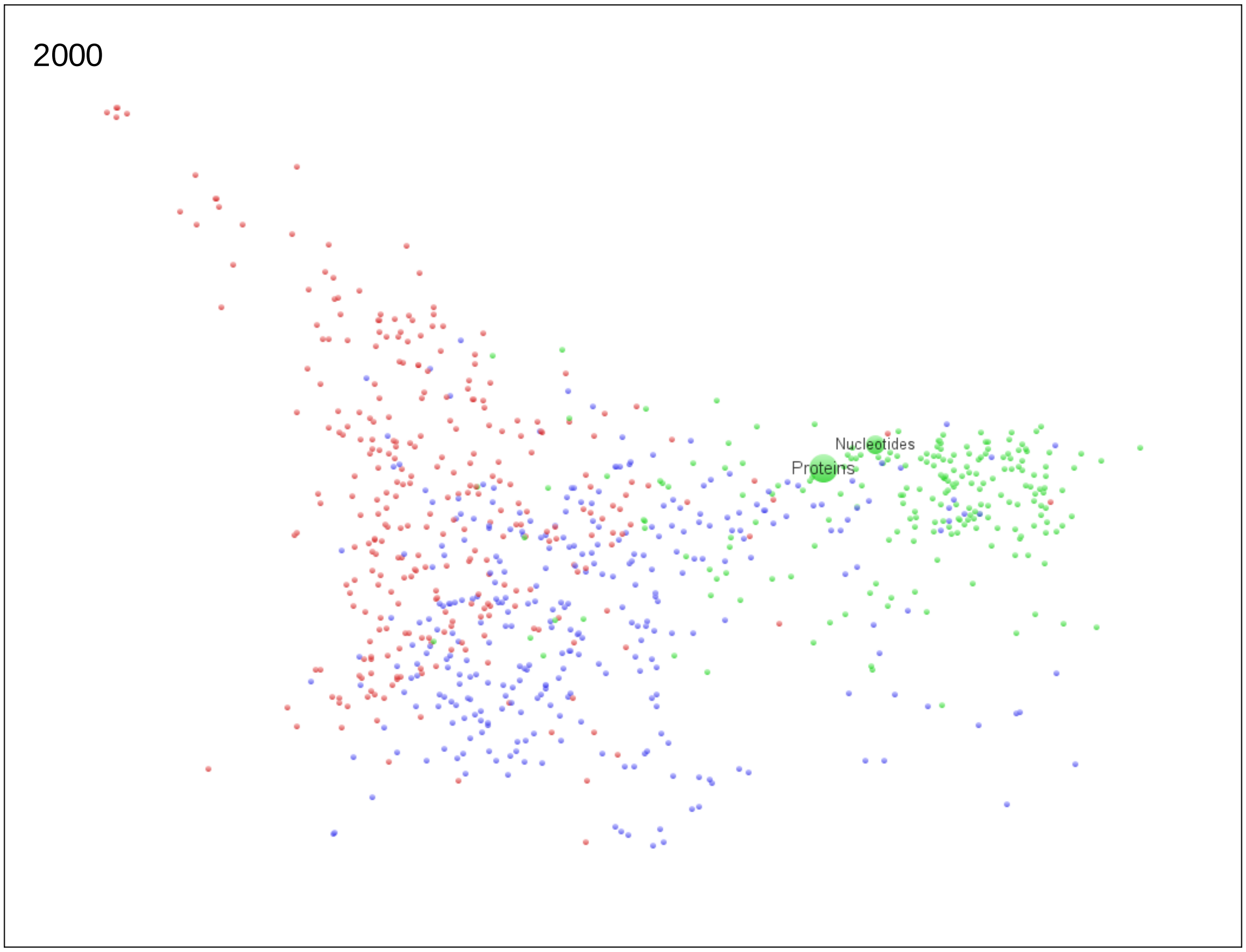

[18] The recall is nine for the year 1997 pre-dating the discovery. These are some records with the words "virus-interfering RNA" in their titles or abstracts.

**Figure 6**: Two MeSH categories used in research about “RNA interference” in 2000, overlayed on a base map of 822 MeSH categories at the first and second level.

Figure 6 shows the overlay for 2000: at this time the research had reached two categories among the second-level categories in the D-branch of “Drugs & Chemicals”: “Proteins” and “Nucleotides”. Perhaps, one can consider this entrance on the map as the beginning of an innovation trajectory in the semantic space spanned at this level of the index, whereas before this year the orientation was towards analysis, and indexed at more specific—that is, deeper—levels of the index tree. The year thereafter (2001, not shown here), “Genetic Techniques” was added as a major label to the set, but the MeSH term “Nucleotides” was no longer involved, notwithstanding that RNA is composed of nucleotides. In other words, the representation has entered into a next stage: it is still laboratory-confined, but techniques and equipment (functionalities) are added to the question of the biochemical composition that was generalized at this index level before that time. As noted, the representation of the research process at lower levels was probably different.

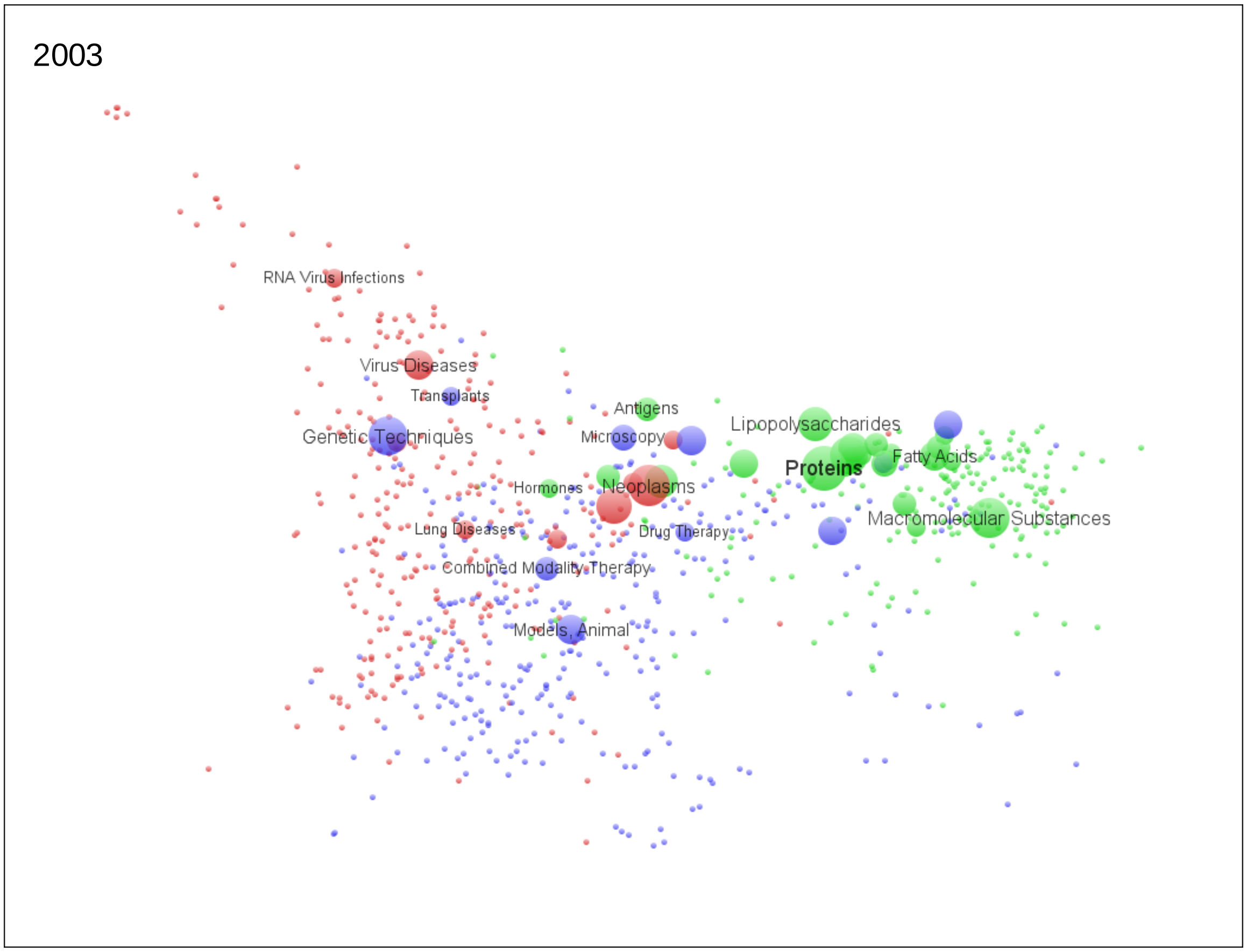

**Figure 7:** Overlay of 42 MeSH used in 2003 for mapping "RNA interference" on the base map of 822 MeSH (Diseases, Drugs, and Techniques map), using VOSviewer.

In the year thereafter (2002; not shown here), further development reaches the clinical domain. Leydesdorff & Rafols (2011) signaled a transition in the dynamics of preferential attachment between 2002 and 2003 as indicative of a change in the diffusion dynamics (Gay, 2010a and b). Figure 7 provides a representation of the situation in 2003 (that is, after this change): all domains of the map are now affected and the new development can in this sense be considered as globalized across medically relevant domains. After 2003, the occupation of the map will grow mainly in terms of the number of categories involved, their size and density.

The full animation for the period 1998-2010 can be retrieved from http://www.leydesdorff.net/pubmed/rnai_vos.pps . Similarly, the maps on the basis of the Pajek baseline can be animated using the file at http://www.leydesdorff.net/pubmed/rnai_paj.pps. As discussed above, the base maps are structurally not so very different between VOSviewer and Pajek. In the Pajek map of recent years (not shown here), the overlapping labels can reasonably be read, but not using VOSviewer if the suppression of overlaps among labels is turned off. The central concepts are more clustered in the center of the MDS-like solution of VOSviewer than in the network overlay of Pajek. In sum, the layout of the two respective algorithms is different, and thus it depends on the research question which system would be optimal for a particular visualization.

In the case of the emergence of "RNA interference", for example, both animations show that "Prognosis" becomes a MeSH term more important than "Genetic Techniques" during the years 2005-2007. Such a global development is visible at a glance using the VOSviewer animation because the main concepts are foregrounded and thus one can see this change emphasized. For a more detailed analysis one may prefer the figures from Pajek, because Pajek additionally allows for the statistical decomposition (in terms of components, etc.). Note that both programs provide the possibility to export the maps into the SVG-format[19] that can be edited and further embellished using Adobe Illustrator or the freeware program InkScape. For example, one can change the colors and sizes of the labels and nodes.

[19] SVG: Scalable Vector Graphics is a markup language for visual representations.

## 7. Further extensions

### *7.1. Other levels of the MeSH hierarchy*

The choice of the first two levels of the hierarchy above was made for pragmatic reasons because this cut provided us with a domain with sufficient (822) categories for a relevant visualization. Using one next-deeper level of the index would have added 6,907 categories, of which 4,237 in the three branches ("C", "D", and "E") under study. Another possibility, however, would be to collapse all categories in levels three or deeper into the corresponding category at the second level. "Bacterial endocarditis" in the category class of C14.260.249, for example, would thus be counted as an instance of "Cardiovascular infections" as its super-class C14.260. Many more and more fine-grained co-classifications are thus induced in the same maps. Would this enable us to represent the research process in more detail than by focusing only on second-level categories?

We tested this option at the level of both the base maps and overlays. A base map which is based on the collapsing of the lower-level categories into the second-level categories in the three branches ("C", "D", and "E") is, on the one hand, more densely packed when compared with the base maps provided in Figures 3 and 4 above for Pajek and VOSviewer, respectively. The overlap among the three main colors is also larger. (One would be able to counteract upon this visual effect technically by using a higher threshold for the cosine.) However, the factor structure of the data matrix is almost similar to the one without this incorporation of lower-level categories, whereas the visualizations are more difficult to interpret than the ones provided in Figure 3 and Figure 4.

The overlays including the lower-level categories, on the other hand, are so much richer than the ones provided above that we decided to make this routine and the corresponding database additionally available as an alternative at http://www.leydesdorff.net/pubmed/pubplus.exe and http://www.leydesdorff.net/pubmed/pubplus.dbf. The routine ("pubplus.exe") generates the overlays otherwise similarly as vos.txt and pajek.vec, but after subsuming the lower-level categories under the second-level ones. (In this case, the 117 first-level categories are ignored.)

This more richly informed representation of the environment in terms of co-classifications provided us with an unexpected finding from the perspective of our research question. For example, the map for RNA Interference in 2000 provided in Figure 6 can be compared with the new map for 2000 (Figure 8). This latter map seems to inform us about the situation in *later* years. Can one consider perhaps Figure 8 as the specification of an expectation in 2000 for the map in 2003 (Figure 7)? Can the latent dimensions of a new technology thus be made visible in earlier years?

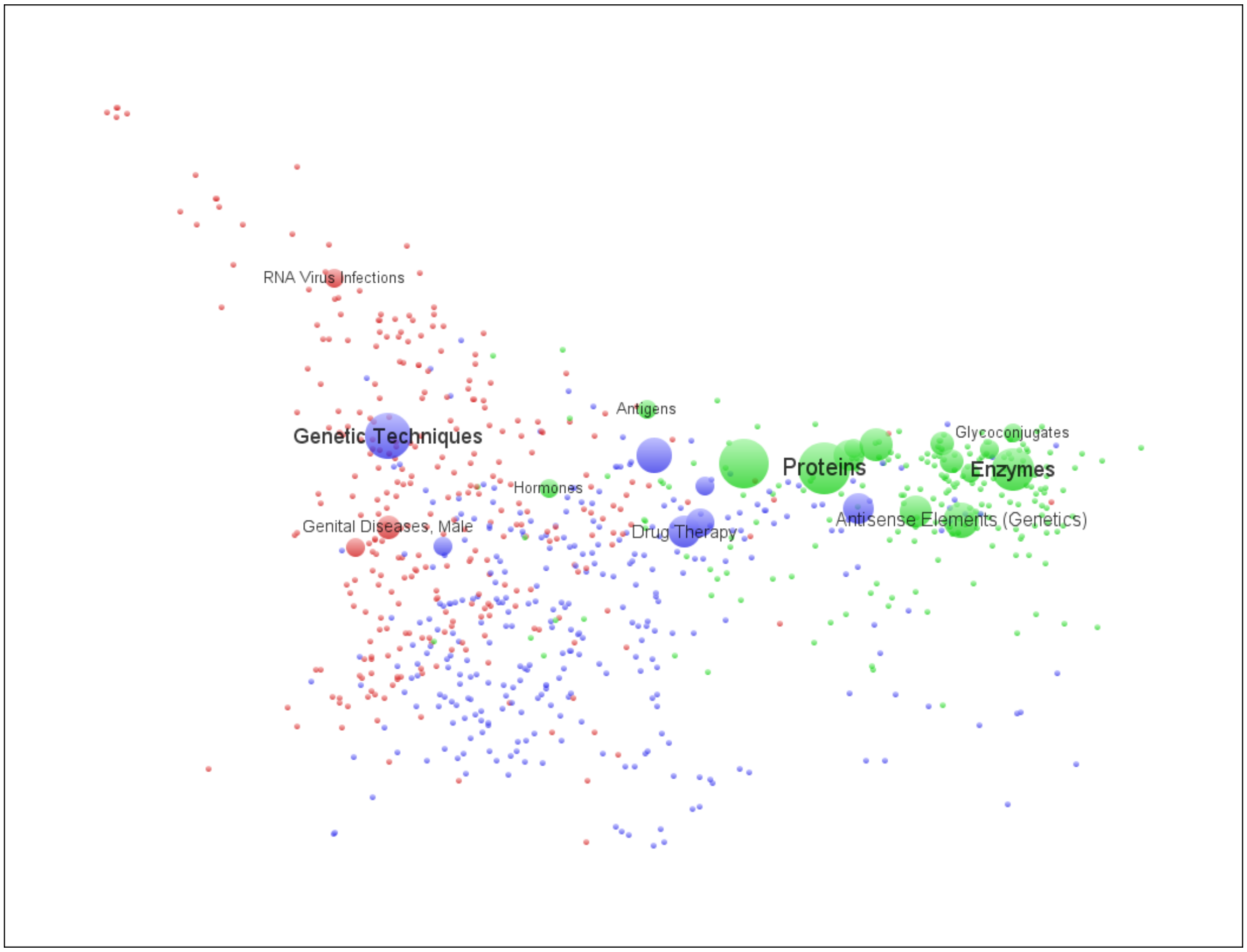

**Figure 8**: Overlay of 25 MeSH terms used in 2000 in RNA interference; MeSH terms provided at levels deeper than two are aggregated at the second-level.

In the case of RNA interference, 486 MeSH categories occurring 515 times in the set of 112 documents with publication year 2000 (see Table 4) were collapsed into 25 MeSH categories at the second level (Figure 8). In 2003, however, 1197 (that is, ten times as many) documents show a structure (in Figure 7) that can be compared with this expectation in 2000. In other words, the relevant environments may already be visible in an earlier year, but the finer-grained distinctions in the scholarly literature reach the higher levels in the tree structure of the index only in a later year. These dynamics are reminiscent of the dynamics between "restricted" and "elaborate" discourses which can operate in parallel, but also change over time (Bernstein, 1971; Coser, 1975).

In other words, developments at lower levels of the index tree can be expected to follow a dynamics different from those at higher levels: knowledge content generated at the research front has to be packaged to be made an "immutable mobile" accessible for specialists in other disciplines (Latour, 1987, at pp. 226f.). Within specialties the emphasis can be on deconstruction, analysis, and precision, but the translation process requires a certain extent of black-boxing and standardization when one wishes to communicate across disciplinary and institutional boundaries. These changes in the dynamics can also be considered as selection processes in different selection environments (Leydesdorff & Rafols, 2011). Since selection is deterministic, specific patterns can be expected to emerge in these production-diffusion dynamics. Thus, different "signatures" of techno-scientific developments in databases can be expected (Leydesdorff *et al.*, in preparation).

### *7.2. Interfaces with other databases*

The Medline/PubMed database is publicly available and therefore mirrored at different locations. From the bibliometric perspective, the interfacing with the two main bibliometric databases—WoS and Scopus—is most relevant since these two databases contain citation information. Scopus provides PubMed Identification Numbers (PMID) among the output fields and the WoS is integrated with Medline in the Web-of-Knowledge of Thomson Reuters. As noted, the output of this latter version of Medline can be read directly by the routines described above by choosing option 2 at the opening screen. (In addition to the input files for the mapping, the routine organizes the text file into a relational database.)

The address information in Medline is less standardized than in the WoS or Scopus, and is mostly limited to corresponding authors. A specific routine would be required for interfacing this information with geo-codes needed for overlays to Google Maps. The output files of Scopus and/or WoS containing citation information can also be used, but one needs an additional step. For this purpose, we added the generation of one more file to the already existing routine "ISI.exe" (for organizing output files of the *Science Citation Index*; available at http://www.leydesdorff.net/software/isi/index.htm).

The new file ("batch.txt") is formatted so that it can be cut-and-pasted into the so-called NCBI[20] Batch Citation Matcher of PubMed at http://www.ncbi.nlm.nih.gov/pubmed/batchcitmatch. The output of this matching program is returned to one's email address. When saved at "match.txt," this file can be used as input to a routine "match.exe" available at http://www.leydesdorff.net/pubmed/match.exe. The output of this routine called "pmid.txt" can be used as input at the advanced search engine of PubMed (discussed above).[21] Thus, recalls from the *Science Citation Index* (WoS) can selectively be used as input to the overlays developed for PubMed.

As noted, Scopus output already contains the PMID numbers. (Scopus can also be searched directly with PMIDs.) When exported as a comma-separated file from Scopus, one can copy the column with the heading "PubMed ID" from Excel to a text file to be named "match.txt";

[20] The National Center for Biotechnology Information (NCBI) advances science and health by providing access to biomedical and genomic information; it is part of the U.S. National Institute of Health.

[21] Additionally, the routine match.exe writes the PMID numbers into the WoS output file ("core.dbf") for relational database management if this file is available in the same folder.

“match.exe” can also read this file and will recognize the header as a file from Scopus. The resulting output file “pmid.txt” is equivalent to the one generated in the case of WoS data. Whether one wishes to use Scopus or WoS depends on the research question and the institutional availability of the database(s).

In future, we would be interested in developing a routine in the opposite direction, namely to use MeSH terms for the citation analysis and perhaps the normalization (Bornmann *et al.*, 2008; Leydesdorff & Bornmann, 2011). As noted, the MeSH terms may provide a delineation of the reference sets that is more informed at the article level than (groupings of) journals. In principle, such an interface is feasible since the citation data was coupled to the Medline data within the WoS since the introduction of the WoS v5 interface in July 2011. However, this will be the subject of another paper (Leydesdorff & Opthof, in preparation).

## 8. Conclusions

The main focus of the present paper may seem technical and methodological, but our argument is also theoretical. However, we wish to develop both perspectives—the technical of software development and the theoretical of innovation studies—in relation to each other. A central issue in innovation studies has been to relate scientific and technological developments to clinical applications, or in the terminology of Agarwal & Searle (2009): “demand” and “supply.”

Hitherto, science-oriented databases such as the *Science Citation Index* and practitioner-oriented ones like PubMed have been developed in parallel, and this has hindered the study of the nonlinear dynamics of interactions between demand and supply that are basic to knowledge-based innovations. A first step in breaking down this barrier is to develop instruments that allow us to provide the descriptive statistics of the relevant data within a single analytical framework. The base maps enable us to visualize the complex dynamics in the data from different perspectives.

How does a new science-based technology diffuse after its emergence? Studying the dynamics of social networks, Leydesdorff & Rafols (2011; cf. Gay, 2010a and b) proposed a model of two stages which they associated with Schumpeter Mark I (Schumpeter, 1912) and Schumpeter Mark II (Schumpeter, 1943), but then not in the business environment, but in the realm of research (cf. Freeman & Soete, 1997). Unlike the mode-1/mode-2 distinction as a transition at the systems level (Gibbons *et al.*, 1994), our focus remains on the evolution of specific technologies.

In the initial stage (Mark I), the dynamics is disciplinarily organized in terms of the specialties involved in the invention. New entrants to the network preferentially attach in this stage to the inventors. At a later stage, the invention becomes packaged as an innovation so that its contents can move across disciplinary borders and thus be used in other specialties. At this stage, a division of labour among "centres of excellence" in metropolitan cities (such as Boston, London, Seoul, etc.) is generated and the preferential attachment (Barabási & Albert, 1999; Leydesdorff & Rafols, 2011; Newman, 2004) is increasingly focused on relating to authors in these centers.

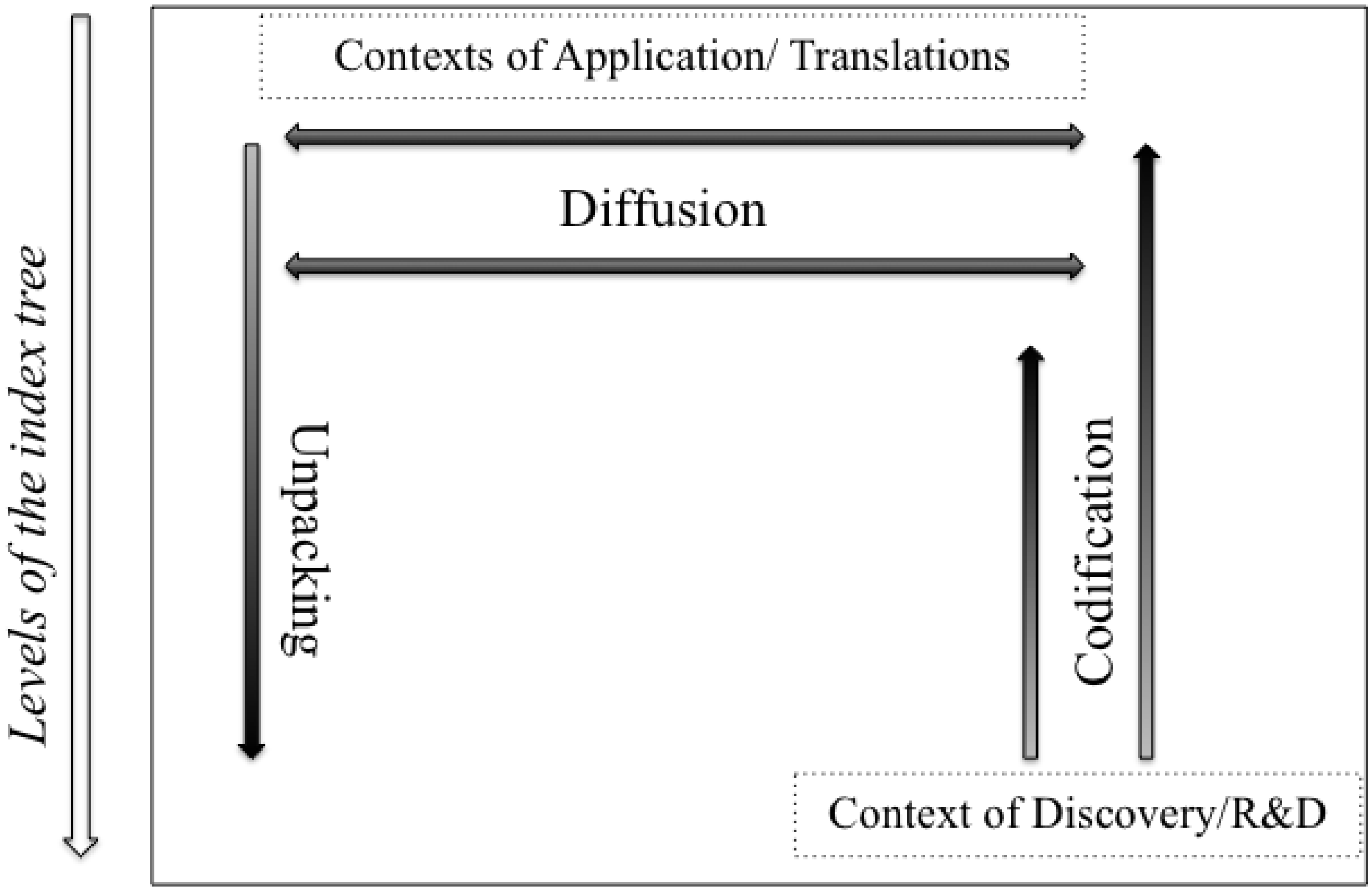

**Figure 9**: Two stages of permeation into the database in terms of levels of an index tree.

Soete & Ter Weel (1999) suggest "creative accumulation" as a label for this stage as against Schumpeter (Mark I)'s well-known "creative destruction" needed in the early stages of the emergence of a new technology. The technology in the later stage becomes globalized (as a regime), while in the previous stage it developed along local trajectories. In other words, the technology is both transformed by and transforms its contexts (Figure 9). One can expect that the carrying scholars themselves are aware of these changing selection environments, and thus adjust their preferences in the respective attachments competitively.

We used above also Latour's (1987) characterization of this double dynamics as "immutable mobiles." On the one hand, the content of the "mobiles" has to be constructed in processes of R&D. On the other, these contents have to be made stable ("immutable") to the transfer process across interfaces. The two configurations provide different selection environments that can be expected to operate concurrently and interact. However, the nonlinear dynamics of selection environments operating upon each other can be expected to lead to specific trajectories that leave traces in the databases.

From this perspective, the well-known reaction-diffusion dynamics may become another relevant metaphor. Stabilization along a trajectory may be meta-stabilized and then globalized into a regime (Geels *et al.*, 2008). The translation map would then be expected to overwrite the production map. However, the map at this higher level of the index (in our case, level 2) may be too coarse for recognition by researchers whose perspective as specialists can only be partial. This raises questions of validation, in our opinion.

Given the different perspectives involved, however, the main question for an overarching bibliometric perspective would be the reliability of visualizations and animations for the study of innovations: can the possible trajectories of medical innovations be classified meaningfully in terms of these nonlinear dynamics? From this perspective, the current study was mainly exploratory. It taught us among other things, that the second-level MeSH terms are too coarse for a representation of the research process itself. The usefulness of the basemap from the perspective of innovation studies needs to be tested further. Thus, the study presented here provides us with an agenda and heuristics for how one can study the nonlinear dynamics of emergent technologies from a bibliometric perspective (Helbing & Balietti, 2011).

**Acknowledgements**
We acknowledge support by the ESRC project 'Mapping the Dynamics of Emergent Technologies' (RES-360-25-0076). Ismael Rafols acknowledges funding from US National Science Foundation (Award #0830207, "Measuring and Tracking Research Knowledge Integration"). The findings and observations contained in this paper are those of the authors and do not necessarily reflect the views of the funding agencies. The authors are grateful to Lutz Bornmann and Tobias Opthof for comments on previous drafts.